\documentclass[conference]{IEEEtran}
\usepackage{tikz}
\usepackage{fancyvrb}
\usepackage{framed}

\usepackage{tcolorbox}
\tcbuselibrary{skins,breakable}
\usepackage{listings}
\usepackage{xcolor}
\usepackage{caption}
\usepackage{multirow}

\usetikzlibrary{arrows.meta, positioning, shapes.geometric}

\newtcolorbox{grayroundbox}{
  breakable,
  colback=gray!10,
  colframe=gray!60,
  arc=4mm,
  boxrule=0.6pt,
  left=10pt,right=10pt,top=8pt,bottom=8pt,
  enhanced
}

\definecolor{bgStruct}{RGB}{235,242,255}   
\definecolor{bgFlat}{RGB}{235,247,235}     

\newcommand{\hlStruct}[1]{\colorbox{bgStruct}{\strut #1}}
\newcommand{\hlFlat}[1]{\colorbox{bgFlat}{\strut #1}}

\definecolor{codeborder}{RGB}{140,140,140}
\definecolor{codenumber}{RGB}{150,150,150}
\definecolor{codekeyword}{RGB}{0,90,160}
\definecolor{codecomment}{RGB}{120,120,120}

\lstdefinestyle{cleanC}{
  backgroundcolor=\color{white},
  frame=single,
  rulecolor=\color{codeborder},
  frameround=tttt,
  basicstyle=\ttfamily\footnotesize,
  numbers=left,
  numberstyle=\tiny\color{codenumber},
  numbersep=8pt,
  xleftmargin=18pt,        
  framexleftmargin=0pt,   
  framexrightmargin=6pt,
  framextopmargin=4pt,
  framexbottommargin=4pt,
  keywordstyle=\color{codekeyword},
  commentstyle=\color{codecomment}\itshape,
  stringstyle=\color{black},
  showstringspaces=false,
  tabsize=2,
  breaklines=true,
  breakatwhitespace=true,
  captionpos=b,
  language=C,
  lineskip=-2pt
}

\usepackage{graphicx}   
\usepackage{caption}    
\usepackage{enumitem}
\IEEEoverridecommandlockouts
\usepackage{cite}
\usepackage{amsmath,amssymb,amsfonts}
\usepackage{algorithmic}
\usepackage{graphicx}
\usepackage{textcomp}
\usepackage{xcolor}
\usepackage[a4paper, total={184mm,239mm}]{geometry}

\usepackage{threeparttable}

\begin{document}

\title{LLM4PQC - Accurate and Efficient Synthesis of PQC Cores by Feedback-Driven LLMs}


\author{
\IEEEauthorblockN{%
Buddhi Perera\IEEEauthorrefmark{1}\IEEEauthorrefmark{2},
Zeng Wang\IEEEauthorrefmark{1}\IEEEauthorrefmark{2}\IEEEauthorrefmark{3},
Weihua Xiao\IEEEauthorrefmark{1}\IEEEauthorrefmark{2},
Mohammed Nabeel\IEEEauthorrefmark{2}\IEEEauthorrefmark{3},\\
Ozgur Sinanoglu\IEEEauthorrefmark{3},
Johann Knechtel\IEEEauthorrefmark{3},
Ramesh Karri\IEEEauthorrefmark{2}}
\IEEEauthorblockA{\IEEEauthorrefmark{1}Equal Contributions}
\IEEEauthorblockA{\IEEEauthorrefmark{2}NYU Tandon School of Engineering, New York, USA \quad \IEEEauthorrefmark{3}NYU Abu Dhabi, Abu Dhabi, UAE}
\text{\{ghp7482, zw3464, wzwx2356, mtn2, ozgursin, johann, rkarri\}@nyu.edu}
}

\maketitle
\begin{abstract}
The design of post-quantum cryptography (PQC) hardware is a complex and hierarchical process with many challenges.
A primary bottleneck is the conversion of PQC reference codes from C to high-level synthesis (HLS) specifications, which requires extensive manual refactoring~\cite{basu2019nist,PQCISVLSI,soni2021hardware}.
Another bottleneck is the scalability of synthesis for complex PQC primitives, including number theoretic transform (NTT) accelerators and wide memory interfaces.
While large language models (LLMs) have shown remarkable results for coding in general-purpose languages like Python, coding for hardware design is more challenging; feedback-driven and agentic integration are key
principles of successful state-of-the-art approaches.
Here, we propose LLM4PQC, an LLM-based framework that refactors high-level PQC specifications and reference C codes into HLS-ready and synthesizable C code.
Our framework generates and verifies the resulting RTL code.
For correctness, we leverage a hierarchy of checks, covering fast C compilation and simulation as well as RTL simulation.
Case studies on NIST PQC reference designs demonstrate a reduction in manual effort and accelerated design-space exploration compared to traditional flows.
Overall, LLM4PQC provides a powerful and efficient pathway for synthesizing complex hardware accelerators.
\end{abstract}


\section{Introduction}

As advancements in quantum computers gain momentum, widely used cryptography schemes such as RSA and ECC are at risk~\cite{shor1994algorithms}. To protect critical infrastructure from such
quantum-computer attacks, the National Institute of Standards and Technology (NIST) has started standardizing modern replacements, based on post-quantum cryptography (PQC) algorithms~\cite{NIST_PQC_2024}.
As PQC moves toward broad deployment, there is increasing demand for efficient and trustworthy hardware implementations across platforms, ranging from high-performance servers to energy-efficient embedded devices.

Hardware implementations require longer development cycles than their software counterparts, causing PQC hardware research to lag behind the rapid progress of the cryptographic community.
Traditional hardware development is based on a register-transfer level (RTL) design flow, where architects must explicitly build the hardware architecture and optimize data flow to meet performance and area constraints.
This process demands substantial domain expertise in hardware design, microarchitecture, and low-level optimization, making it time-consuming and difficult to adapt to evolving PQC algorithms.

High-level synthesis (HLS) offers an alternative approach by allowing designers to describe hardware functionality using high-level languages such as C, C++, or SystemC.
HLS significantly reduces the required domain expertise and shortens development time by abstracting architectural and data-flow details. Moreover, HLS enables reuse of reference software implementations provided by the
community, facilitating rapid prototyping and design-space exploration (DSE). While HLS may not achieve the same performance as hand-optimized RTL designs, it does provide an effective trade-off between
productivity and efficiency for accelerating hardware development.

PQC reference implementations are released as portable C code that prioritizes clarity and functional correctness, rather than hardware synthesizability.
Translating PQC C code into HLS implementations requires  manual refactoring and repeated tool-driven iterations~\cite{basu2019nist,PQCISVLSI,soni2021hardware}.
Our work addresses this challenge, advancing hardware implementation efforts and enabling more targeted DSE efforts for PQC algorithms.

We present LLM4PQC, a feedback-driven LLM workflow to transform PQC reference code into synthesizable HLS-C code,
considering end-to-end verification and DSE for power, performance, and/or area (PPA) objectives.
Among other aspects, we focus on refactoring barriers exposed by real-world PQC kernels: HLS-incompatible constructs like calls to \texttt{math.h} functions,
floating-point arithmetic, and runtime routines that build-up constant tables.
Overall, our LLM4PQC framework integrates:
(i) preprocessing transformations tailored to target HLS-C,
(ii) automated C-to-HLS transformation and evaluation (using our C2HLSC~\cite{10.1145/3734524} tool integrated with Catapult for compilation, simulation, and synthesis).
Feedback from compilation/simulation failures
help to iteratively refine the different transformation outcomes.

\textbf{Contributions.} This paper makes three key contributions through LLM4PQC and related experiments:
\begin{enumerate}[leftmargin=*]
  \item LLM-driven preprocessing of C codes for HLS compliance. Particular challenges here are data-structure expansion, dynamic memory allocation, and initialization of constants.
  \item A structured, feedback-driven workflow for transforming PQC reference C codes into synthesizable HLS-C using LLMs, evaluated through an automated and fully integrated
  hardware design loop comprising C2HLSC and Catapult.
  \item An empirical case study on established PQC algorithms (Kyber, Dilithium, and Falcon) and their core primitives.
\end{enumerate}

\textbf{Paper organization.}
Section~\ref{sec:background} reviews key concepts for PQC and HLS, motivating our approach.
Section~\ref{sec:method} details the proposed framework and its components.
Section~\ref{sec:results} presents a detailed empirical case study, and Section~\ref{sec:discussion} discusses limitations and future directions.


\section{Background}
\label{sec:background}

\subsection{PQC Schemes}
In August 2024, NIST finalized its first set of PQC standards: FIPS 203 (Kyber), FIPS 204 (Dilithium), and FIPS 205 (SPHINCS+)~\cite{NISTFIPS203_Kyber, NISTFIPS204_Dilithium, NISTFIPS205_SPHINCS}.
Kyber is a key-encapsulation mechanism (KEM), while Dilithium and SPHINCS+ are digital signature algorithms (DSAs).
Falcon is another DSA selected by NIST, currently under development for standardization as FIPS 206~\cite{FalconENS2018}.
SPHINCS+ is hash-based, whereas the others rely on lattice-based cryptographic hardness assumptions.

\subsection{PQC and Hardware Acceleration}
\label{sec:background:routines}

Certain PQC subroutines are inherently slow when operating on regular CPUs, due to complex operations and irregular memory access patterns. Without dedicated hardware acceleration, these subroutines become performance
bottlenecks.
Next, we discuss key subroutines, which are also summarized in Table~\ref{tab:pqc_subroutines}.


\begin{table}[tb]
\centering
\caption{Complex Subroutines Used in PQC Algorithms}
\label{tab:pqc_subroutines}
\begin{tabular}{lll}
\hline
\textbf{Algorithm} & \textbf{Subroutines} & \textbf{Ref.} \\ \hline \hline
Kyber     & NTT, SHAKE & \cite{aikata2022kali} \\ \hline
Dilithium & NTT, sparse polynomial multiplication, SHAKE & \cite{zhao2024sparse} \\ \hline
Falcon   & Gaussian sampler, NTT, FFT & \cite{ouyang2025falconsign} \\ \hline
\end{tabular}
\end{table}

\begin{figure*}
    \centering
    \includegraphics[trim={0 12cm 0 12cm},clip,width=\textwidth]{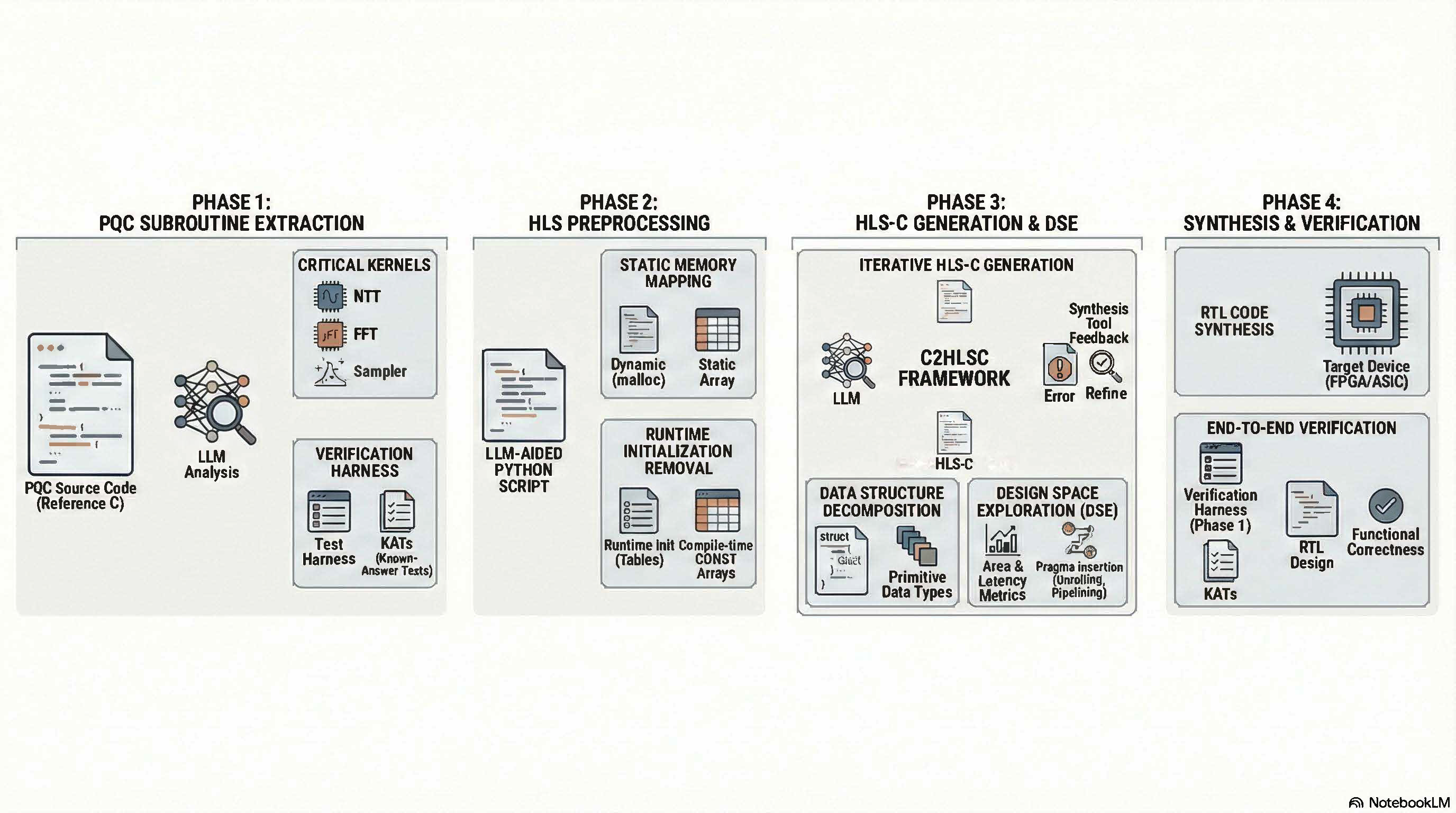}
    \caption{The four phases of LLM4PQC: (i) PQC subroutine extraction, (ii) HLS preprocessing, (iii) HLS-C generation and DSE (based on integrating C2HLSC with Catapult), and (iv) synthesis and verification.
	    Figure generated using NotebookLM.
}
    \label{fig:flow}
\end{figure*}

\subsubsection{Number Theoretic Transform (NTT)}
This subroutine reduces the complexity of polynomial multiplications required in lattice-based schemes from \textit{O($N^2$)} to $\textit{O(N~log~N)}$.
In Kyber and Dilithium, NTT operations dominate the overall execution time~\cite{botros2019memory}.
Falcon utilizes both the Discrete Fourier Transform (DFT) and NTT, and these transform operations together account for a substantial portion of its runtime.

Consequently, accelerating NTT is essential for improving the performance of these schemes. Hardware acceleration is generally preferred over software approaches, as the modular arithmetic and irregular memory access patterns inherent to NTT significantly degrade software performance.

\subsubsection{Sampler}
Sampling (and hashing) operations in lattice-based PQC vary across schemes.
Noise sampling may follow centered binomial distributions (Kyber and Dilithium) or discrete Gaussian distributions (Falcon). For Falcon, sampling alone consumes over 64\% of the
cycles~\cite{ouyang2025falconsign}.

While hardware implementations of binomial samplers are relatively straightforward, often leveraging extendable-output hash functions such as SHAKE, designing an efficient hardware sampler for Falcon's Gaussian
distribution remains considerably more challenging.


\subsubsection{Fast Fourier Transform (FFT)}
The key generation and signature generation phases in Falcon heavily rely on FFT and inverse FFT (IFFT) to enable efficient polynomial multiplication. This is achieved by embedding polynomials from $\mathbb{Z}[x]/(x^n + 1)$ into the complex domain.

Both FFT-based arithmetic and discrete Gaussian sampling require high-precision floating-point arithmetic, which poses significant challenges for efficient hardware implementations.


\subsubsection{Summary and Motivation}
Practical adoption of PQC requires implementations that are not only secure and correct, but also efficient and scalable. Hardware acceleration is essential to meet performance, latency, and energy targets in resource-constrained devices and high-throughput server-side applications. Many PQC candidates use arithmetic kernels such as NTT, FFT, modular arithmetic, sampling, and hash primitives.
These kernels dominate runtime and influence hardware cost.

\subsection{From Reference C to HLS-C: Common Challenges}
\label{sec:background:compliance}

While PQC reference implementations are valuable for correctness and portability, they are not  directly synthesizable.
Converting reference C code into HLS-C requires refactoring software-oriented constructs and making memory/control explicit. We observe the following synthesis blockers:
\begin{itemize}[leftmargin=*]
  \item \textbf{Non-synthesizable math library calls.} Functions from \texttt{math.h} (e.g., \texttt{sin}, \texttt{cos}, \texttt{pow}, \texttt{sqrt}, \texttt{log}, \texttt{exp}) are  unsupported or expensive in HLS flows, motivating synthesizable replacements (e.g., LUT/CORDIC/polynomial approximations) or offline precomputation.
  \item \textbf{Floating-point-heavy arithmetic.} Reference codes often use \texttt{float}/\texttt{double} or floating-like abstractions for accuracy and numerical stability. However, synthesis of such floating-point
  constructs is tool dependent,
  and automated conversion to fixed-point requires careful handling of scaling, rounding, saturation, and constants.
  \item \textbf{Runtime initialization of constant tables.} Cryptographic kernels often precompute twiddle factors, roots of unity, or other constant tables during initialization, e.g., using trigonometric functions.
  Such initialization routines are not synthesizable as is and, thus, must be refactored into compile-time constants or ROM-like tables.
  \item \textbf{Complex data structures and dynamic memory.} Use of \texttt{struct}/\texttt{union} constructs can complicate mapping to hardware interfaces and memory layouts, while dynamic allocation (e.g., using
		  \texttt{malloc}) is not synthesizable and must be replaced with static arrays whose sizes respect tool and hardware constraints.
\end{itemize}

These challenges are found in most if not all reference codes.
For example,
the Falcon sampler is a ``stress test'' for HLS compliance; it comprises complex data structures, dynamic memory allocations, and pointer references in various interfaces.
This motivates us to devise a syntax-aware C transformation/preprocessing stage that preserves correctness while enabling HLS-C synthesis.


\subsection{HLS Compilation, Simulation, and Synthesis}
HLS development is iterative and sequential by nature: small upfront code changes can affect correctness, scheduling, resource usage, and performance; evaluating all these aspects requires to rerun simulation and
synthesis. However, repeated synthesis runs are expensive, which makes naive trial-and-error approaches overly slow.

Our workflow utilizes a staged feedback loop: quick checks via C-level compilation and simulation, to filter out incorrect transformations, followed by HLS synthesis, to obtain PPA metrics for DSE among correct candidates.
Toward automation, our workflow connects LLM-driven code transformation to tool-driven feedback, systematically incorporating error messages and PPA results into subsequent iterations.

\section{Methodology}
\label{sec:method}

LLM4PQC comprises four phases, as shown in Fig.~\ref{fig:flow} and discussed in detail below.

\subsection{PQC Subroutine Extraction}
Converting complete PQC reference implementations to HLS is neither practical nor necessary. While reference implementations prioritize portability and correctness, only specific computational kernels/subroutines warrant
hardware acceleration due to their impact on overall system performance (Sec.~\ref{sec:background:routines}).

We adopt a selective extraction strategy that isolates performance-critical subroutines for HLS conversion while retaining non-critical logic in software.
Our technique employs LLM-aided analysis to systematically identify and extract target subroutines from NIST PQC reference codebases.
We prompt GPT-5.2 to perform three key tasks: (i) identify performance-critical functions, (ii) extract the target function with minimal dependencies while preserving its functional interface, and (iii)
co-generate a verification infrastructure including known-answer tests (KATs), helper functions, and a test harness that enables isolated compilation and functional validation.

This approach ensures that each extracted subroutine can be independently verified at the C level, using reference KATs, before HLS conversion begins.
By maintaining functional equivalence through KAT-based testing, we significantly reduce debugging complexity for subsequent HLS refinement.

\subsection{HLS Preprocessing}

PQC implementations are designed for portability and clarity, rather than synthesis.
We find that passing these implementations directly to C2HLSC~\cite{10.1145/3734524} can fail, despite the fact that C2HLSC has been successfully demonstrated for more generic hardware
domains.
As discussed in Section~\ref{sec:background:compliance}, these PQC codes contain several specific HLS blockers, like
runtime initialization of large constant tables.
Thus, to refactor these codes in a HLS-friendly manner,
we devise preprocessing stages and implement them in an automated LLM-aided pipeline.


\subsubsection{Static Memory Mapping}
To enable C2HLSC to process C code containing dynamic memory allocations,
we
systematically transform dynamic memory usage into static memory allocations.
Specifically, calls to \texttt{malloc} and \texttt{memcpy} in the PQC reference C code are replaced with functionally equivalent static-memory implementations generated using LLMs.
An example prompt is shown below.

\begin{grayroundbox}
\noindent\textbf{Prompt 1: Static Memory Mapping}
\begin{Verbatim}[fontsize=\small]
You are an expert in C code generation. The
following C code contains dynamic memory 
allocation using malloc. Rewrite the code to
eliminate all uses of malloc, using only
static or stack allocation.
If there are any memory frees or functions
that call memoryfrees, remove them as well.
Do not provide main functions or testbenches.
Return only the modified C code. I don't want 
any code explanations.
\end{Verbatim}
\end{grayroundbox}

The transformed C code is then compiled and validated against KATs, ensuring that functional correctness is maintained.
In cases where compilation fails or the test results indicate incorrect behavior, the corresponding error messages are fed back to the LLMs as corrective prompts, to help iteratively refine the generated code.

\subsubsection{Initialization Removal}

This stage targets runtime routines that compute and populate constant tables (e.g., twiddle factors inside FFT algorithms or other lookup tables).
As such routines are not part of the steady-state computation, they may introduce non-synthesizable dependencies in HLS.

This stage follows five steps. First, the LLM is prompted to detect initialization functions that fill global/static arrays.
Second, we prompt the LLM to automatically write a small wrapper/runner program that covers this initialization function and invokes it in a standalone \texttt{main()} routine.
Third, we compile and execute the runner to obtain the fully initialized table values, and we emit them as \texttt{const} array declarations in C syntax.
Fourth, we rewrite the original codebase to remove both the definition of the initialization function and all invocations of it.
Fifth, we insert the generated \texttt{const} array declarations into a header file included by the design, so that the transformed C reads these compile-time constants instead of executing runtime initialization logic,
enabling the HLS tool to readily infer corresponding ROMs during synthesis.

\subsection{HLS-C Generation and Design-Space Exploration}

The preprocessed C code is provided to the LLM-driven C2HLSC framework~\cite{10.1145/3734524} to generate HLS-synthesizable C code.
C2HLSC employs an iterative refinement process in which the input C code is progressively transformed using error feedback from the synthesis tool, augmented by in-context learning examples provided to the LLM.
Each generated HLS-compatible C version is validated for functional correctness by comparing its outputs against those produced by the reference C implementation, based on the KATs.

PQC reference implementations frequently rely on complex user-defined data structures, which hindered C2HLSC's ability to produce correct synthesizable code.
To address this limitation, we extended the C2HLSC framework with structure-aware prompting mechanisms as follows.
When synthesis errors related to user-defined structures were detected, customized prompts instruct the LLM to decompose structured data types into equivalent representations composed exclusively of primitive data types.
An example prompt is shown below.

\begin{grayroundbox}
\noindent\textbf{Prompt 2: Data Structure Expansion}
\begin{Verbatim}[fontsize=\small]
Update the code to eliminate the use of 
structs. To eliminate the use of structs, 
unpack their fields into plain variables 
within the top function and use the unpacked
variables as arguments to top_function_hls.
\end{Verbatim}
\end{grayroundbox}

This transformation enabled synthesis while preserving functional equivalence with the original reference code. Note that this is implemented within C2HLSC rather than as another preprocessing stage, given
the complexity and diversity of different user-defined structures, which requires more context-focused operation than standalone preprocessing could offer.

To optimize the synthesizable C code with respect to area and latency, we perform DSE through an additional LLM-aided optimization loop within our core component (C2HLSC and Catapult).
Here, the LLM was prompted to introduce HLS pragmas, specifically for loop unrolling and pipelining. The optimization process was guided by quantitative feedback from the synthesis tool, including area utilization and latency metrics obtained from previous synthesis runs. This feedback-driven exploration enabled the identification of design configurations that best satisfied the specified optimization objectives.


\subsection{Synthesis and Verification}

The generated code can be readily (re)synthesized for any hardware target of choice.
We do so for both ASIC and FPGA settings in an extensive case study, as discussed next.
Based on the KATs generated for the C code, using Catapult, we also automatically generate testbenches for the synthesized RTL.
This ensures end-to-end verification of correct behaviour.

\section{Case Studies}
\label{sec:results}

\subsection{Experimental Setup}

\textbf{Benchmark Selection and Extraction}: We evaluate LLM4PQC on five representative PQC subroutines: Kyber-NTT, Dilithium-NTT, Falcon-NTT, Falcon-Sampler, and Falcon-FFT.
Using our LLM-aided extraction, we obtain standalone C codes from NIST reference implementations along with verification infrastructure (KAT tests and test harnesses).

\textbf{Preprocessing and HLS-C Conversion}:
Following the workflow in Fig. 1, extracted C code undergoes preprocessing (Sec.~III-B) to eliminate HLS blockers including dynamic memory allocation, runtime initialization, and complex data structures.
The preprocessed code is then fed into C2HLSC, which iteratively refines the code through LLM-driven optimization with feedback from Catapult HLS compilation and simulation until functional correctness is verified, along
with DSE for PPA objectives (area, latency).
Each benchmark underwent 10 independent runs to account for LLM non-determinism.

\textbf{LLM Model}:
Without loss of generality, we evaluate LLM4PQC on \textit{ChatGPT o3-mini} with these settings: 0.2 temperature, 0.2 nucleus sampling and 4096 maximum tokens.

\textbf{Synthesis and Evaluation}:
A run is considered as passing if: (i) HLS-C code compiles successfully in \textit{Catapult HLS 2023.1\_2/1049935}, (ii) KAT-based simulation verifies correctness, and (iii) timing closure is achieved.
We report success rate and average/min/max metrics from passing runs.
Passing designs undergo RTL synthesis for ASICs using \textit{Design Compiler} targeting the \textit{Nangate 45nm} cell library.
For FPGA synthesis, we use \textit{Vivado} targeting \textit{Xilinx Artix-7} (Kyber/Dilithium) and \textit{XCZU7EV} (Falcon).

\textbf{RTL Simulation and Verification}:
The design under test (DUT) is the RTL output from Catapult, while the testbench itself is C-based and leverages the simulator's programmable logic interface (PLI) to apply stimuli and observe responses.
The same KAT stimuli are applied to the C model for HLS, and its outputs are compared against the DUT responses.
All designs are validated using 1,000 test cases / KATs.
Simulations are performed using \textit{ModelSim SE-64 version 2021.3}.

\textbf{Metrics}:
We evaluate runs across two key dimensions with related metric: (i) \textit{automation efficiency} (success rate, compile runs, and HLS runs), and (ii) \textit{hardware performance} (area in LUTs/FFs/DSPs/BRAMs, latency in
		cycles/µs).

\begin{figure}[t]
\centering
\begin{lstlisting}[style=cleanC,
escapeinside={(*@}{@*)},
caption={Sampler function from Falcon reference code, with user-defined data structures and other HLS-incompatible constructs highlighted.}]
int sampler(void *ctx, fpr mu, fpr isigma)
{
(*@\hlStruct{   \ \ sampler\_context *spc;}@*)
    int s;
    fpr r, dss, ccs;
(*@\hlStruct{    \ \ spc = ctx;}@*)
    s = (int)fpr_floor(mu);
    r = fpr_sub(mu, fpr_of(s));
    dss = fpr_half(fpr_sqr(isigma));
(*@\hlStruct{    \ \ ccs = fpr\_mul(isigma, spc->sigma\_min);} @*)
    for (;;){
        int z0, z, b;
        fpr x;
(*@\hlStruct{        \ \ \ \ \ \ z0 = gaussian0\_sampler(\&spc->p);} @*)
(*@\hlStruct{        \ \ \ \ \ \ b = (int)prng\_get\_u8(\&spc->p) \& 1;} @*)
        z = b + ((b << 1) - 1) * z0;
        x = fpr_mul(fpr_sqr(fpr_sub(fpr_of(z), r)), dss);
        x = fpr_sub(x,
            fpr_mul(fpr_of(z0 * z0), fpr_inv_2sqrsigma0));
(*@\hlStruct{        \ \ if (BerExp(\&spc->p, x, ccs))} @*)
        {
            return s + z;
        }
    }
}
\end{lstlisting}
\vspace{-5mm}
\end{figure}


\begin{figure}[t]
\centering
\captionsetup{type=lstlisting}
\begin{lstlisting}[style=cleanC,
escapeinside={(*@}{@*)},
caption={Generated, HLS-friendly Falcon sampler function with flattened arguments and structures highlighted.}]
(*@\hlFlat{int sampler\_hls(fpr mu, fpr isigma,}@*)
(*@\hlFlat{             \ \ \ \ \ \ \   uint8\_t p\_buf[512],}@*)
(*@\hlFlat{             \ \ \ \ \ \ \   size\_t p\_ptr[1],}@*)
(*@\hlFlat{             \ \ \ \ \ \ \   uint8\_t p\_state[256],}@*)
(*@\hlFlat{             \ \ \ \ \ \ \   fpr sigma\_min)}@*)
{
    int s, z0, z, b;
    fpr r, dss, ccs, x;

    s = (int)fpr_floor(mu);
    r = fpr_sub(mu, fpr_of(s));
    dss = fpr_half(fpr_sqr(isigma));
(*@\hlFlat{    \ \ ccs = fpr\_mul(isigma, sigma\_min);} @*)
    for (;;)
    {
(*@\hlFlat{    \ \ \ \ \ \ z0 = gaussian0\_sampler\_no\_structs(p\_buf,}@*) 
(*@\hlFlat{    \ \ \ \ \ \ \ \ \ \ \ \ \ \ \ \ \ \ \ \ \ \ \ \ \ \ p\_ptr, p\_state);} @*)
(*@\hlFlat{    \ \ \ \ \ \ b = ((int)prng\_get\_u8\_no\_structs(p\_buf,} @*)
(*@\hlFlat{    \ \ \ \ \ \ \ \ \ \ \ \ \ \ \ \ \ \ \ \ \ \ \ \ \ \ p\_ptr, p\_state)) \& 1;} @*)
        z = b + (((b << 1) - 1) * z0);
        x = fpr_mul(fpr_sqr(
                fpr_sub(fpr_of(z), r)), dss);
        x = fpr_sub(x,
            fpr_mul(fpr_of(z0 * z0),
                    fpr_inv_2sqrsigma0));
(*@\hlFlat{      \ \ \ \ \ \  if (BerExp\_no\_structs(p\_buf, p\_ptr,} @*)
(*@\hlFlat{    \ \ \ \ \ \ \ \ \ \ \ \ \ \ \ \ \ \ \ \ \ \ \ \ \ \ p\_state, x, ccs))} @*)
            return s + z;
    }
}

int sampler(void *ctx, fpr mu, fpr isigma)
{
(*@\hlFlat{    sampler\_context *spc = (sampler\_context *)ctx;} @*)
(*@\hlFlat{    return sampler\_hls(mu, isigma,} @*)
(*@\hlFlat{    spc->p.buf.d, \&spc->p.ptr, spc->p.state.d,} @*) 
(*@\hlFlat{    \ \ \ \ spc->sigma\_min);} @*)
}
\end{lstlisting}
\vspace{-5mm}
\end{figure}

\subsection{Preprocessing for HLS Compliance}
As indicated, the sampler and FFT subroutines required preprocessing.
For the sampler, dynamic memory allocations were refactored into fixed-size static arrays whose dimensions are known at compile time, enabling compatibility with static analysis and synthesis flows.
For FFT, runtime array initializations were substituted with constant lookup tables to eliminate dynamic initialization overhead.
The NTT subroutines did not require any code changes during preprocessing.

\subsection{HLS-C Conversion}
During HLS-C generation using our integrated C2HLSC and Catapult flow, we find that the sampler subroutines required substantial modifications, due to
the expansion of user-defined data structures into primitive data types as required by C2HLSC.
Accordingly, modifications were applied not only to the top-level function but also to all invoked subfunctions.
To ensure correct data-structure expansion while preserving the original top-level interface required by the main testbench, a wrapper function was introduced, as illustrated in Listing 1 and 2.
Importantly, all these modifications were fully assisted by the proposed LLM-driven workflow.

Other subroutines required mostly minor adjustments, primarily involving the replacement of pointer-based accesses with array-based representations already supported by C2HLSC.
In addition to these structural changes, HLS pragmas were introduced to enable DSE and performance optimization. These pragmas guided Catapult to apply loop unrolling and pipelining to selected subroutines.



\begin{table*}[htb]
\centering
\caption{Success rate, runs for C compilation and HLS, and synthesis results. HLS Synthesis is configured for area optimization. Results are for the \textit{Nangate 45nm} ASIC library.}
\label{tab:hls_results}
\begin{tabular}{l c c c c c c c c c c c c c}
\hline
{\textbf{Benchmark}} &
{\textbf{Succ.}} &
\multicolumn{3}{c}{\textbf{\# Compile Runs}} &
\multicolumn{3}{c}{\textbf{\# HLS Runs}} &
\multicolumn{3}{c}{\textbf{Area [$\mu$m$^2$]}} \\

& \textbf{[\%]} & \textbf{Avg.} & \textbf{Min.} & \textbf{Max.} & \textbf{Avg.} & \textbf{Min.} & \textbf{Max.} & \textbf{Avg.} & \textbf{Min.} & \textbf{Max.} \\
\hline
\hline
Kyber-NTT       &   100 & 14.22 &  10   & 22    & 7.67  & 6     & 11    & 2,957.54   & 2,781.7    & 3,048.5  \\ 
Dilithium-NTT   &   100   &   11.56   & 10     & 16     & 6.44     & 6     & 9     & 6,506.94         & 6,137.1         & 6,822.2       \\
Falcon-NTT      & 70     & 13.17     & 10     & 17     & 6.00     & 1     & 11    & 162,689.64         & 160,646.1         & 163,163.2    \\
Falcon-Sampler  & 40    & 15.75    & 12    & 20    & 7.25   & 6     & 8     & 141,465.1 & 93,961.07   & 159,425.27    \\
Falcon-FFT      & 100     & 11.67     & 10     & 18     & 6.67     & 6     & 10     & 56,802.5         & 44,702.2         & 62,360.5 \\
\hline
\end{tabular}
\end{table*}

\begin{table*}[htb]
\centering
\begin{threeparttable}
\caption{Performance and resources comparison with related work for FPGA implementations.}
\label{tab:fpga_result}
\begin{tabular}{l l c r r r r r r r}
\hline
\textbf{Primitive} & \textbf{Work}   & \textbf{FPGA}  & \textbf{Freq ($MHz$)} & \textbf{\# Cycles}  & \textbf{LUTs} & \textbf{FFs} & \textbf{DSPs} & \textbf{BRAMs} & \textbf{Latency ($\mu s$)} 
\\ \hline\hline                                                                                                                        

\multirow{2}{*}{Kyber-NTT}       & \cite{nguyen2024high}        & Artix-7       & 250         & 277         & 4,834        & 4,683         & 0           & 1          &  1.11  \\
                & LLM4PQC                                       & Artix-7       &  75         & 5,639        & 146         & 119          & 3           & 1          & 75.00  \\ \hline
                                                                                                                                           
\multirow{2}{*}{Dilithium-NTT}   & \cite{nguyen2024high}        & Artix-7       & 180         & 128         & 7,451        & 5,275         & 0           & 0          & 0.71 \\
                & LLM4PQC                                       & Artix-7       &  70         & 6,664        & 370         & 258          & 5           & 2          & 95.20   \\ \hline
                     
\multirow{2}{*}{Falcon-NTT\textsuperscript{*}}      
                & \cite{mu2022scalablefalcon}                   & Artix-7       & 117         & 4,468        & 2,119        & 1,058         & 8           & 3          & 38.00      \\
                & LLM4PQC                                       & Artix-7       &  38         & 30,741       & 982         & 555          & 7           & 2          & 809.00     \\ \hline
                     
\multirow{2}{*}{Falcon-FFT\textsuperscript{\#}}
                & \cite{dam2025falcon}                          & Artix-7       & 134         & 2,048        & 17,395       & 7,950         & 20          & 4          & 15.30   \\
                & LLM4PQC                                       & Artix-7       &  25         & 10,768       & 8,447        & 1,945         & 33          & 2          & 430.00    \\ \hline
                                                                                                                                           
\multirow{2}{*}{Falcon-Sampler\textsuperscript{\dag}}  
                & \cite{Cornett2025DeusExLLMs}                  & XCZU7EV       & 300         & 111         & 7,683        & 8,266         & 108         & 0          & 0.73  \\
                & LLM4PQC                                       & XCZU7EV       & 150         &  59         & 17,619       & 3,927         & 18          & 1          & 0.39  \\ \hline 

\end{tabular}

\begin{tablenotes}
\footnotesize
\item[\textsuperscript{*},\textsuperscript{\#}] 
Falcon-NTT assumes a polynomial degree of 1024; Falcon-FFT assumes a polynomial degree of 512.
\item[\textsuperscript{\dag}] 
\cite{Cornett2025DeusExLLMs} uses LLM-assisted RTL generation via HLS. Their results are based on ChatGPT o4-mini, whereas ours are based on o3-mini.
\end{tablenotes}

\end{threeparttable}
\end{table*}

\subsection{HLS Performance: Area and Latency}

For the ASIC setting, representing a baseline obtained through Catapult, Table~\ref{tab:hls_results} reveals considerable variation across the primitives optimized for area.
Kyber-NTT produces the most compact designs at 2,957.54 $\mu$m² on average,
while Dilithium-NTT requires approximately 2.2× that area at 6,506.94 $\mu$m².
The Falcon primitives demand even more resources:
Falcon-FFT averages at 56,802.5 $\mu$m² and Falcon-NTT averages at 162,689.64 $\mu$m², respectively, which is due to their hybrid FFT/NTT architectures and floating-point support.
Furthermore, the Falcon-Sampler exhibits significant variability (93,961.07–159,425.27 $\mu$m²), indicating a diverse range of LLM-driven architectural settings, ranging from compact sequential to aggressive parallel implementations.


Latency extracted from the synthesis reports provide unreliable cycle counts, which are thus not reported.
This is because several loops in the primitives' algorithms are not statically bounded; their iteration counts cannot be correctly determined at compile time by the HLS tools.
Although the actual cycle counts are not precise, their ranges and orders of magnitude still provide valuable insights. For NTT and FFT,
the variation in cycle counts is small across all primitives. For the Falcon-Sampler, however, varations are significant. This is caused by the data-dependent nature of Gaussian rejection sampling, where lower latencies correspond to aggressive optimizations, while higher values represent conservative implementations that preserve the full iterative logic.

\subsection{RTL Simulation and Verification}

Table~\ref{tab:fpga_result} compares the PPA of our FPGA designs against related work.
Most baselines rely on manual RTL tailored to specific microarchitectures, with the exception of the sampler in~\cite{Cornett2025DeusExLLMs}, which also utilizes an LLM-to-HLS flow.
Our generated sampler outperforms~\cite{Cornett2025DeusExLLMs} in latency. Furthermore, for Kyber, Dilithium, and Falcon NTT primitives, LLM4PQC achieves lower area utilization (LUTs and FFs) compared to prior art.

\section{Discussion}
\label{sec:discussion}

Our experience with LLM4PQC reveals several key insights for the automated hardware synthesis of cryptographic primitives. One of the most significant challenges is the conversion of floating-point operations,
particularly evident in Falcon-FFT. The primary obstacle is not numerical accuracy, but rather that general-purpose LLMs lack the specific HLS domain knowledge to translate floating-point heavy implementations into
synthesizable hardware. Consequently, generated code often compiles successfully as standard C but fails during synthesis due to unsupported library functions or improper resource mapping. This necessitates our
preprocessing phase, which transforms reference code at the C level to remove these blockers prior to LLM conversion. Furthermore, we find that starting with cleaner, well-structured implementations such as those
from the PQC-clean repository~\cite{SSR:KSSW22} significantly reduces efforts compared to using original NIST submissions.

We observed that LLM4PQC tends to generate highly area-efficient designs, often surpassing manual baselines in LUT and FF utilization. However, this efficiency comes at the cost of higher latency. This suggests that, without explicit guidance, the LLM favors compact, sequential loop structures over the aggressive unrolling and
parallelization strategies typical of hand-crafted cores. This variance confirms that coding style directly dictates hardware quality and reveals substantial opportunities for DSE at the C
level. Shifting the framework's focus from fundamental synthesizability to performance tuning, specifically targeting latency reduction through architectural refinements, remains a key frontier.

\section{Conclusion}

We have presented LLM4PQC, a novel framework for translation of PQC reference implementations into synthesizable hardware accelerators.
By integrating LLM-driven refactoring with a compile-simulate-synthesize feedback loop, we successfully generated functional hardware for complex kernels, including the Falcon Sampler and NTT/FFT primitives.
Our experimental results demonstrate that LLM4PQC effectively navigates the trade-off between design productivity and hardware quality. Our designs prove highly competitive, achieving lower resource utilization than
several hand-optimized manual baselines. While this compactness currently incurs a latency penalty compared to manual designs, our approach outperforms the state-of-the-art LLM-based prior
work~\cite{Cornett2025DeusExLLMs} in both FF utilization and latency. This confirms that a feedback-driven workflow is superior to direct LLM translation.

Promising directions for future work include closer/agentic integration of the LLM-driven phases in LLM4PQC, fine-tuning the LLMs on HLS datasets, or employing retrieval-augmented generation (RAG) to query
verified HLS patterns during code generation. 
Additionally, we envision an LLM-driven evolutionary framework where we iteratively mutate and explore C codes for ease of hardware generation (e.g., via loop restructuring or array partitioning). 
To accelerate such exploration, pre-synthesis PPA prediction models, similar to VeriLoC~\cite{hemadri2025veriloc}, could provide rapid estimates for timing and area without the latency of full synthesis runs.

\section{Acknowledgments}

Buddhi Perera and Ramesh Karri were supported in part by the National Science Foundation under Grant No. \#2450539.
The authors acknowledge support from the NYU Center for Cybersecurity (CCS) and the NYU Abu Dhabi Center for Cybersecurity (CCS-NYUAD).

\IEEEtriggeratref{11}
\bibliographystyle{IEEEtran}
\bibliography{references}

\end{document}